\documentclass[12pt]{article}

\usepackage{scicite}
\usepackage{url}
\usepackage{lineno}
\usepackage{graphicx}
\usepackage{amsmath}
\usepackage{tcolorbox}
\usepackage{times}

\newenvironment{sciabstract}{%
\begin{quote} \bf}
{\end{quote}}

\newcounter{lastnote}

\usepackage[a4paper, margin=1in]{geometry}  % Adjust margins as needed
\usepackage{authblk}

\title{Moving towards informative and actionable social media research}

\author{
Joseph B. Bak-Coleman,$^{1,2,3,4,5, 6\ast}$ Stephan Lewandowsky,$^{7,8}$ 

Philipp Lorenz-Spreen,$^{9,10\ast}$ Arvind Narayanan,$^{11,12}$ 

Amy Orben,$^{13}$ Lisa Oswald$^{14, 10}$ \\

\normalsize{$^{1}$ Centre for the Advanced Study of Collective Behavior, University of Konstanz, Konstanz, Germany} \\
\normalsize{$^{2}$ Department of Collective Behavior, Max Planck Institute of Animal Behavior, Konstanz, Germany} \\
\normalsize{$^{3}$ Department of Biology, University of Konstanz, Konstanz, Germany} \\
\normalsize{$^{4}$ Santa Fe Institute, Santa Fe, NM, USA} \\
\normalsize{$^{5}$ Berkman Klein Center, Harvard University, Cambridge, MA, USA} \\
\normalsize{$^{6}$ Department of Biology, University of Washington, Seattle, WA, USA} \\
\normalsize{$^{7}$ School of Psychological Science, University of Bristol, Bristol, UK} \\
\normalsize{$^{8}$ Department of Psychology, University of Potsdam, Potsdam, Germany} \\
\normalsize{$^{9}$ Center Synergy of Systems and Center for Scalable Data Analytics and Artificial Intelligence, Dresden University of Technology, Dresden, Germany} \\
\normalsize{$^{10}$ Center for Adaptive Rationality, Max Planck Institute for Human Development, Berlin, Germany} \\
\normalsize{$^{11}$ Center for Information Technology Policy, Princeton University, Princeton, NJ, USA} \\
\normalsize{$^{12}$ Department of Computer Science, Princeton University, Princeton, NJ, USA} \\
\normalsize{$^{13}$ MRC Cognition and Brain Sciences Unit, University of Cambridge, Cambridge, UK} \\
\normalsize{$^{14}$ Center for Critical Computational Studies (C3S), Goethe-University, Frankfurt am Main, Germany}
\vspace{25px}
\\\normalsize{The order of authors is alphabetically}\\
\normalsize{$^\ast$To whom correspondence should be addressed; E-mail: jbakcoleman@gmail.com, philipp.lorenz-spreen@tu-dresden.de}
}

\date{}

\begin{document} 

% Double-space the manuscript.

\baselineskip24pt

% Make the title.

\maketitle

% Place your abstract within the special {sciabstract} environment.

\begin{sciabstract}
Social media is nearly ubiquitous in modern life, raising concerns about its societal impacts---from mental health and polarization to violence and democratic disruption. Yet research on its causal effects is still inconclusive: Various methods, spanning observational to experimental, can yield seemingly conflicting results. Considering the complexity of such socio-technical systems, with coupled networks, feedback loops and collective phenomena, this may not be surprising. Here, we enumerate and examine the features of social media as a complex system that challenge our ability to infer causality at societal scales. Attempts to ascertain and summarize causal effects have tended to prioritize findings from randomized controlled trials (RCTs). However, like observational studies, RCTs rely on assumptions that may frequently be violated in the context of social media, especially regarding societal outcomes at scale. Drawing on insight from disciplines that have faced similar challenges, like climate-science or epidemiology, we propose a path forward that combines the strengths of observational and experimental approaches while acknowledging the limitations of each. Progress, we argue, requires moving beyond isolated, linear effects to mechanistic explanations of how social media platforms generate collective outcomes.

\end{sciabstract}

\section*{Introduction}

Social media have permeated most societies, impacting citizens' lives around the world. Nearly 5 billion people across the globe are using social media, and penetration rates exceed 80\% in northern Europe, 70\% in North-, 67\% in South-America, and reach 75\% in Eastern Asia \cite{statista2024}. The defining attribute of social media platforms is that users not only consume but also create and distribute content. In addition, social media are characterized by some degree of algorithmic curation and in most cases also an advertisement-driven business model \cite{Lewandowsky22}. 

Accompanying the prominence of social media is widespread concern about their adverse effects on society, ranging from undermining public health measures \cite{Allington20} and mental health \cite{orben2022windows, valkenburg2022social} to exacerbating polarization \cite{kubin2021role}, fomenting extremism \cite{Kursuncu19}, and disrupting democracy \cite{Gerbaudo23}. A plethora of research has sought to address those concerns; Google Scholar returns more than 1,500,000 publications since 2020 in response to the query ``social media'' (Search on 30 March 2026). At least tacitly, this substantial research endeavor is concerned with generating a consensus \textit{causal} understanding of the relationships between social media use and the outcomes of interest. Absent such an understanding, any necessary corrective action, from design to regulation, may remain ineffective. 

Unfortunately, most efforts trying to understand the impact of social media on large-scale societal changes --- to ultimately judge whether they warrant corrective action --- are compelling enough to raise concern but fail to yield a scientific consensus. 
Broadly speaking, observational evidence has uncovered associations between social media use and political participation, polarization, trust in political institutions, body image, and well-being \cite{lorenz2023systematic, orben2024mechanisms, valkenburg2022social, teague_digital_2026}. 

Aiming to gain causal insight into these associations, some observational work has leveraged quasi-experimental approaches \cite{angrist1996identification} to estimate the causal impact on trust \cite{miner2015unintended, sabatini2019online}, political participation \cite{enikolopov2020social, geraci2022broadband}, hate crimes \cite{bursztyn2019social, muller2021fanning}, or populist support and polarization \cite{lelkes2017hostile, schaub2020voter}, as well as mental health  \cite{braghieri2022social} (for a more detailed overview, see \cite{lorenz2023systematic}). 

Further studies aim at causality by running Randomized Controlled Trials (RCTs) that modify some aspect of users' experience with social media. A wide range of such experiments have been carried out, ranging from following specific accounts or installing experience-altering browser extensions to news-feed modification or abstention from platforms altogether \cite{bail2018exposure, levy2021social, allcott2020welfare, asimovic2021testing, allcott2024effects, guess2021consequences, kelly2024web, lyngs2024finally, guess2023reshares, guess2023social, nyhan2023like, beknazar2025toxic, piccardi_reranking_2025, gauthier_political_2026}. 

Despite this considerable effort to gain causal insights, effects are still scattered across various outcome measures and largely varying in their size, which led to widely different interpretations in scientific and public debate. For outcomes such as polarization, extremism, and voting patterns, the estimates in the literature tend to be the least clear \cite{national2023social}. Often the resolution to seemingly conflicting observational and experimental work has been to assume or assert that RCTs provide better, less-biased causal estimates. For example, high-profile experiments conducted in conjunction with Meta were argued to ``rule out'' moderate sized or larger effects of their platforms on various societal outcomes such as extremism and polarization \cite{nyhan2023like, allcott2024effects,guess2023reshares}. Similarly, a review of the impacts of misinformation cited experiments as providing causal evidence for the absence of social media's effects on polarization, which was later argued to obviate concerns stemming from observations, qualitative work, and journalistic investigations \cite{Budak2024}. In the context of adolescent mental health, an authoritative report by the National Academies declared that longitudinal evidence could not be used to establish causality, implying formal experiments were necessary and sufficient \cite{national2023social}. 

In this article, we critically examine those claims. We show that they may arise from a misunderstanding that RCTs are necessary and sufficient for establishing causality regarding societal impacts of social media. 
We highlight how the defining feature of social media---social interactions, often facilitated through algorithms \cite{friemel_public_2023}---can produce all manner of feedback, emergent system-level properties \cite{Bayer20}, and temporal dynamics (see Fig.~\ref{fig:sketch} for an overview). Those are properties that constitute ``complex systems'' \cite{anderson1972more, ladyman2013complex}, in which it is the interactions of their parts that give rise to collective behavior that cannot be understood from the analysis of individual parts alone, thereby complicating causal inference. For example, they can undermine the assumptions required by RCTs and likely lead to downward-biased effect sizes, a general challenge recently reviewed in the social sciences more broadly by Bailey et al. \cite{bailey2024causal}. Additionally, we discuss how, when, and whether the estimates generated by RCTs meaningfully correspond to the causal effects of social media on society and how they need to be combined with observational and theoretical approaches if not. Overall, we argue that it is important that interpretations take the scale of the units of analysis into account, without assuming linear extrapolation, neither from the individual to the collective-level nor from the collective to the individual.

However, even if we resolve long-standing debates about social media's net historical impact on society, it may leave us far from the kind of scientific insight needed to guide intervention by policymakers and technologists. To overcome these barriers, we draw on insights from other scientific disciplines, such as climate change and public health, where feedbacks and interactions also frustrate the standalone use of tools such as RCTs. Following in their footsteps, we highlight a path forward for social media research that embraces the strengths of a wide variety of approaches while appreciating and offsetting their limitations. 
Beyond benefits for uncovering mechanism instead of average, linear effects, we argue this approach will enable research to focus on questions that reveal more mechanistic, actionable insight into the relationships between platform design, use, and societal phenomena.

\section*{Estimating social media effects}

We cannot simultaneously observe the presence and absence of a treatment on a single unit (e.g., person, society, etc.). There is no way to directly measure whether an individual or society would be better or worse off with social media. This is true in science more broadly and known as the  ``fundamental problem of causal inference'' \cite{Rubin1974}. To work around this problem, scientists instead rely on relating some source of variation to an outcome of interest. For observational studies, the source of variation is naturally occurring, whereas experiments artificially induce variation in the exposure to treatment. Often, this variation is induced through randomization into treatment and control (i.e., an RCT), thereby externally varying exposure to the presumed causal variable of interest while controlling the effect of other factors through randomization.

Observational approaches to inferring causality have proliferated in past decades, following the so-called ``credibility revolution'' rooted in econometrics and subsequently diffused across the social sciences \cite{angrist_credibility_2010, imbens_causal_2015}. These design-based approaches exploit naturally occurring variation to draw causal inferences. For instance, natural variation in the roll-out of Facebook across college campuses has been used as source of variation to draw inferences about its impacts on mental health \cite{braghieri2022social}. Similarly, variation in smartphone policies in schools have been used to examine impacts on educational and mental health outcomes \cite{Abrahamsson2024}. Outages and feature roll-outs provide additional sources of variation to be exploited. Often, these sources of variation are framed as ``instrumental variables'' which impact treatment exposure but are not believed to impact outcome \cite{angrist_identification_1995}.
Any of these observational approaches to causality require specific \textit{assumptions} to hold for unbiased causal estimation. Non-random roll-outs, changes in the algorithm or smartphone bans, for instance, could confound the primary treatment effect of interest. Addressing these challenges requires adding co-variates to the statistical model to adjust for sources of bias and ideally identify the causal pathway of interest \cite{hernan_causal_2020}.

To avoid these complexities, scientist routinely opt to instead induce variation by randomly assigning treatments to units. At least in expectation, the randomization into treatments and controls addresses confounding and yields an unbiased causal estimate \cite{Deaton18}. Yet this feature often leads to the misunderstanding that RCTs intrinsically yield causal inference, when they too draw causal inference based on assumptions about the nature of, and relationships between, measured values \cite{Cook18c}. 

For example, differential attrition between treatment and control groups is a commonly encountered pitfall to the estimation of unbiased average treatment effects. 
To circumvent this, researchers resort to estimating conditional treatment effects or include a variety of co-variates to the statistical model. Social media research faces these and other challenges to identifying causal effects using RCTs that are underappreciated in the literature, some of which we discuss in the following.

\subsection*{Collective vs. individual-level effects}

First and foremost, it is essential that any empirical estimate of a causal effect corresponds to the causal effect of interest (i.e., the theoretical estimand \cite{lundberg_what_2021}). Many areas of concern regard fundamentally collective phenomena, such as polarization, spread of misinformation, or even aspects of democratic backsliding.
By collective, we mean that a phenomenon of interest realizes across groups of individuals via their interactions, rather than individuals in isolation. For example, affective  polarization can only be meaningfully measured by contrasting feelings towards in-group and out-group parties and therefore considering the relational dynamic that defines polarization \cite{iyengar_origins_2019}. Beyond individual platforms, those collective outcomes can further extend across multiple platforms \cite{mekacher_systemic_2023} and political systems \cite{lorenz2023systematic}. %For example, while mental health can be studied as an individual phenomenon, its trends and changes over time are implicitly societal phenomena that are of interest. 

\begin{figure}[!htb]
    \centering
    \includegraphics[width=0.9\textwidth]{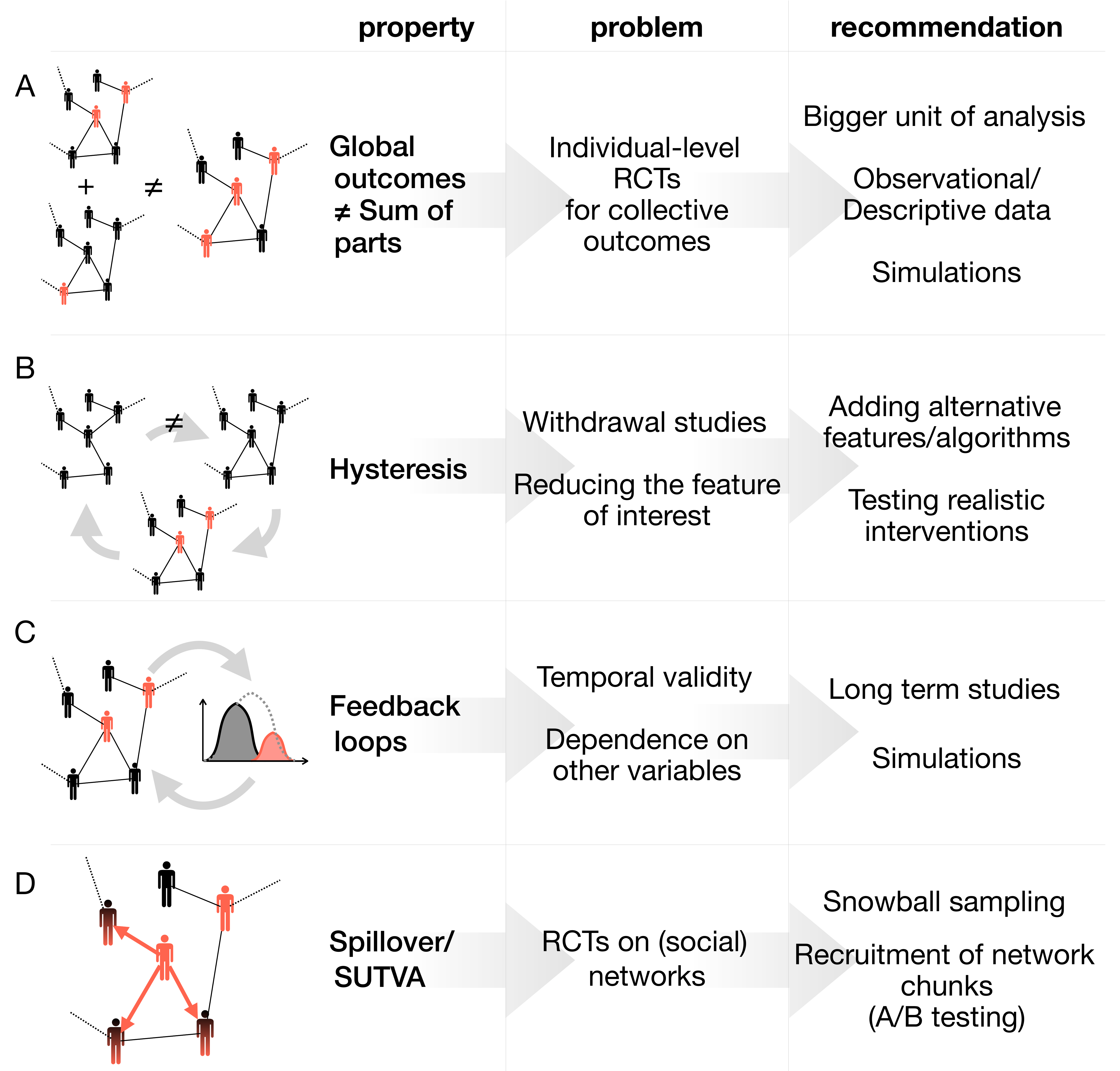}
    \caption{Illustrations of key properties of complex systems that make some RCTs particularly difficult to interpret. \textbf{A} Global outcomes do not linearly depend on the sum of their individual components, making conclusions from small groups to larger societies generally difficult. \textbf{B} More specifically, the outcomes depend on the history of the system, known as \emph{hysteresis}, which makes conclusions about any reverse effects impossible. \textbf{C} Feedback loops that prevent conclusions about one variable without considering the value of another at a specific point in time. \textbf{D} Spillover effects in network-structures, known as
    violations of 
    the ``stable unit treatment value assumption'' (SUTVA),  prevent a clean separation of treatment and control conditions.}
    \label{fig:sketch}
\end{figure}

\begin{tcolorbox}[title=Example A: Collective outcomes,
title filled=false,
colback=blue!5!white,
colframe=blue!75]
As a concrete example, a nation-wide experiment on Facebook in the U.S. involved a treatment whereby users could share that they had voted and see how many others had done the same \cite{bond201261}. The direct, individual treatment effect yielded 60,000 additional votes compared to a control group which did not see social information. However, the authors went one step further and examined social contagion effects via their Facebook friends, which led to an additional 280,000 votes. In this case, the societal estimand is more than five times larger than the individual estimand. Positive feedback, such as the amplification of voter turnout, is a common feature of complex systems and can frustrate estimation of societal effects from individual outcomes. Those findings were later replicated during the 2012 U.S. presidential election \cite{jones_social_2017}. Negative feedback, by contrast, can dampen such effects or even potentially reverse their direction. Often, the societal effect will depend on a mix of positive and negative feedback, making it very difficult to infer the societal effect from individual effects alone. 
\end{tcolorbox}

Distinctions of scale are rare in prominent experiments on social media's impacts, which often take the form of randomizing individuals into different conditions to induce variation \cite{nyhan2023like, allcott2024effects, guess2023reshares}. Yet in doing so, the estimand implicitly but inescapably changes from the effect on collective phenomena to the effect on individuals. Taking the estimate from such a study to indicate a collective effect requires an often unstated assumption---that collective effects over longer periods of time (which is often the estimand of interest) are equal to the simple sum of individual effect on shorter time scales (the estimated quantity, Fig. \ref{fig:sketch}A). Claims about the past effect of a platform on some outcome additionally require assumptions that the direction and magnitude of an effect is time and scale-invariant; from the birth of the platform with its first user to the time-frame of the experiment \cite{baumgartner_stabilization_2025}. 

Unfortunately, causal effects on society will rarely be identical to the effects on individuals. Social media is defined by interactions between individuals, a hallmark of \textbf{complex} systems more broadly. Although the term ``complex'' takes on many meanings across contexts, here we are referring to systems in which interactions between individual parts yield \textit{emergent} system-level properties \cite{anderson1972more, mitchell2009complexity}. A defining feature of emergent properties in complex systems is that they cannot be inferred from the simple sum of the system's constituent parts (cf.,``More is Different'' \cite{anderson1972more, mitchell2009complexity}) (Fig. \ref{fig:sketch}A). This overarching property of complex systems has some, more specific downstream consequences that we discuss in the following and that are relevant for RCTs on and about social media.

\subsection*{Hysteresis}

The ubiquity of social media makes it essentially impossible to find a sample of non-users that is representative of a, for example, general national population of interest, as a control group. Owing to this and potential ethical concerns if expecting negative impact of use, withdrawal experiments that reduce access to social media platforms or platform features are often used \cite{allcott2020welfare, asimovic2021testing, arceneaux2024facebook, bursztyn2025non}. 

\begin{tcolorbox}[title=Example B: Hysteresis,
title filled=false,
colback=blue!5!white,
colframe=blue!75]
To illustrate, suppose we want to evaluate the causal effect of Duolingo (a language learning platform) on language fluency, and we do so by randomly selecting long-time existing users to discontinue learning for a week. We are bound to find that users largely retain their fluency after a week of withdrawal of the platform. 

\medskip
Of course, it would be incorrect to assume the minimal decline in fluency reflects minimal causal effect of Duolingo on language skills accumulated over years of learning. Here, the confound of past use has resulted in downward bias of the estimated effect size. Moreover, for anything that is learned faster (or slower) than it is forgotten, the estimate will be further biased accordingly. 
\end{tcolorbox}

In general, there is no reason to believe that changes in beliefs, behaviors, attitudes (e.g., trust in institutions), or alterations of mental health acquired over years of time spent on a social media platform will rapidly or symmetrically revert upon brief (e.g., weeks or months) withdrawal, if such reversals are even possible or likely (also see Fig. \ref{fig:sketch}B). 

Similar considerations apply to other social outcomes: For example, mental health might transition from one stable state (healthy) to another (unhealthy) because of social media, but reverting to a healthy state is not simply an subtractive feature of ``undoing'' the original trigger \cite{scheffer2024dynamical}. Similar arguments have been made about the stabilization of media effects through repeated exposure and hence non-trivial reversibility \cite{baumgartner_stabilization_2025}. This lack of reliable symmetry between the effect of a cause and its reversion is a common feature of complex systems known as hysteresis. 
Just as we might not expect individuals to return to a prior state, there is little reason to expect that societal effects of social media will revert as platform use, features, and affordances change. It is conceivable, for instance, that a change in societal phenomena becomes self-reinforcing such that the platform's causal effect is akin to a match on dry tinder. Once the fire is burning, the effect of the match would become difficult to identify. As such, hysteresis presents challenges for both observational and experimental research alike.  

\subsection*{Feedback loops}

The third problem associated with interpreting individual-level RCTs is that the effects of interest may unfold non-monotonically over time (Figure~\ref{fig:sketch}C) \cite{klingelhoefer_possible_2026, vandenbosch_digital_2025}. Feedback loops are a well-established concept, describing situations in which two entities iteratively influence one another over time, often reinforcing a particular trajectory. For example, the feedback loop of addiction means that having a cigarette can lead to a relief (e.g., reduced tension or stress) in a long-term smoker. When such long-term smokers are asked to withdraw from smoking for two weeks, they are likely to become irritable (and gain weight, among other side effects of withdrawal). The clear health benefits of smoking cessation would only be observable over much larger time-scales. Under the reasoning common in social media studies involving
RCTs, short-term studies of smoking cessation would incorrectly conclude that smoking is a safe and effective means of losing weight and reducing irritability (also see Fig. \ref{fig:sketch}C). Likewise, relevant outcomes on social media may exhibit temporal dynamism so we should anticipate that findings across experiments of differing lengths will conflict and short-term results, while most prevalent in the field for practical reasons, may be of very limited value to extrapolate to the long term. 

\begin{tcolorbox}[title=Example C: Feedback loops,
title filled=false,
colback=blue!5!white,
colframe=blue!75]
In the context of social media, feedback loops are especially pronounced, largely due to the presence of learning algorithms. For instance, YouTube’s video recommendation system and a user’s preferences may form such a feedback loop. On the one hand, the algorithm aims to infer the user’s preferences from observed behavior. On the other hand, repeated exposure to specific recommendations can lead users to adjust their preferences relative to those recommendations. The algorithm subsequently updates its inferences based on these shifted preferences, thereby perpetuating the cycle. Consequently, measuring user decisions relative to algorithmic recommendations at a single point in time is insufficient to disentangle the historical dynamics that gave rise to the observed state \cite{lewandowsky2024challenges}.
\end{tcolorbox}

\subsection*{Spillover}

Finally, a core reason why RCTs are limited when aiming to study effects of social media is that the individual units interact (Figure~\ref{fig:sketch}D). As described by Eckles et al. ``In online social networks, the behavior of a single user explicitly and by design affects the experiences of other users in the network.'' \cite{Eckles17}

Formally, the problem here is that any RCT would violate a dimension of the ``stable unit  treatment value assumption'' (SUTVA)---or non interference---which requires that the treatment outcome of one experimental unit (e.g., a participant) is not affected by the treatment status of other units \cite{Imbens24, Eckles17, aridor_experiments_2024}. 

On the level of analysis there is a body of literature that aims to account and test for spillover issues in networks \cite{Eckles17, athey_exact_2018}. This is a non-trivial endeavor and requires precise data about network structure and about social connections beyond the treatment, to estimate the extend of spillover. In other words, to distinguish the relative contributions of direct- and indirect effects on the total effect \cite{athey_exact_2018}. Recent work formalizes causal estimands under interference, distinguishing between direct, indirect, and total effects \cite{savje_average_2021}. In networked settings, standard estimators such as differences in means may recover mixtures of these effects, complicating interpretation. This is particularly relevant when the downstream effects of an intervention are of central interest (e.g., voting decisions \cite{jones_social_2017}). Importantly, even when such quantities are well-defined, they need not correspond to collective, system-level outcomes of interest, which are the primary focus of our argument. 

\begin{tcolorbox}[title=Example D: Spillover/SUTVA,
title filled=false,
colback=blue!5!white,
colframe=blue!75]
People are coupled through a myriad of on- and off-platform connections. An individual randomly exposed to an experimentally-altered newsfeed, for example by switching to a chronological sorting, will nonetheless experience algorithmic curation: When neighbors in their network share content they saw through algorithmic curation, the individual in the chronological condition is still indirectly subject to algorithmic sorting.

\medskip
Similarly, the mental well-being of a teenager taken off social media will not just depend on that experimental manipulation, but on whether their friends or parents also stop using these platforms. It follows that violations of SUTVA can be generally expected to minimize the impact of any intervention (such as social media restriction or adaptation), biasing results. 
\end{tcolorbox}

Ultimately, to assess collective outcomes while avoiding violations of SUTVA, the experimental design itself needs to be adjusted: Entire networks, at least subnetworks, need to be randomized into treatment and control states, also known as ``graph cluster randomization'' \cite{ugander_graph_2013}). That is, all friends and family on social media must be assigned to the same treatment (or control) as the focal person of interest whose behaviour is recorded. This procedure can be more or less feasible or effective, depending on the underlying network and its community structure \cite{Eckles17}. %To our knowledge, no existing RCTs purportedly studying societal effects of social media fulfill that criterion. 
Far from hypothetical, these issues have been known to social media companies for decades and exactly such strategies have been developed to address them in their own A/B testing of product features \cite{ugander_graph_2013, karrer_network_2021, gui2015network}. Nonetheless, industry-academic collaborations studying societal effects have relied on individual randomization \cite{guess2023reshares, nyhan2023like}. 

A common theme emerges from the challenges of applying RCTs to infer collective causal effects. Small numbers of individual effects can nonetheless lead to large societal phenomena, as demonstrated in the example for collective outcomes of voting behavior on Facebook described in the first info-box above. Hysteresis too, would be anticipated to generally produce downward bias---especially when past use has stabilized media effects \cite{baumgartner_stabilization_2025}. Positive feedback loops would similarly be anticipated to yield experimental under-estimation when effect are measured on shorter time-scales than the feedback takes to manifest. Finally, spillover and SUTVA violations will tend towards systematic underestimation on social media \cite{karrer_network_2021}. 

Although overestimation is certainly possible, the general expectation from the features above would be systematic downward bias (i.e., towards zero) of effect sizes (see also \cite{cunningham_how_2023} for a discussion). We provide an example of overestimation later in the piece, whereby the effect of misinformation interventions is likely over-estimated and other examples are certainly conceivable. Nevertheless, the frequent null and near-zero effects observed in social media RCTs may often be an artifact of downward bias rather than true absence of meaningful effects---an interpretation we read nevertheless \cite{nyhan2023like}. Unfortunately, as many of these issues are implicit to study design, there is no trivial way to \textit{post hoc} adjust for such bias. 

\subsection*{Interpreting RCTs on social media}

Despite the characteristics of complex systems that make the causal analysis of the societal impacts of social media particularly challenging, various types of RCTs have recently been conducted to investigate precisely these questions. We discuss specific aspects and limitations of these experiments in light of the characteristics described above.

A series of prominent examples is the collaboration between Meta and independent researchers attempting to bridge the gap in our causal understanding of how algorithmic curation affects political outcomes. In the experiments, some users were exposed to a number of interventions, such as receiving less politically-aligned content in their news feeds \cite{Nyhan23}, experiencing a reduced  number of direct reshares from social contacts \cite{guess2023reshares}, being switched to a chronologically sorted feed \cite{guess2023social}, or deactivating the entire platform for 6 weeks \cite{allcott2024effects}. 

Generally speaking, these interventions altered users' exposure to content. For example, default platform features (e.g., algorithmic sorting) tended to expose users to more content that was ideologically aligned and uncivil but also less moderate sources, political content and more uncivil content and like-minded sources compared to a reverse-chronological feed \cite{guess2023social}. The studies additionally conducted surveys to evaluate downstream effects on relevant attitudes, beliefs and behaviors. Here, effects were generally smaller, less consistent, and more ambiguous. For example, reducing exposure to content from like-minded sources did not decrease affective polarization or ideological extremity \cite{Nyhan23}. The authors asserted that their estimated (null) causal effects on individuals provided insight into the societal-scale estimand \cite{Nyhan23}.
%At first glance, these findings are consistent with some concerns that engagement-optimized social media exacerbates polarization and fuels hate via increased exposure to uncivil and like-minded content. However,  
\\
\\
We can reappraise those and other studies that employ RCTs on social media in light of the issues surrounding complexity raised above:

\begin{itemize}

\item First, there is little reason to believe that changes in collective state (e.g., polarization) from increased platform use or platform-feature 
exposure can be directly inferred from changes in exposure for an ensemble of individuals. Such changes involve not only the direct effects of exposure (or withdrawal), but also the changes following exposure to the dynamics across individuals
(global outcomes, see Fig.~\ref{fig:sketch}A). 

\item Moreover, platform-polarized individuals may not revert to a more moderate state merely because exposure to the platform or content is reduced (hysteresis, see Fig. \ref{fig:sketch}B). This applies not just to individuals, but likewise to the structure of the network. In the case of polarization, for example, the network of co-partisan individuals one follows could be heavily shaped by algorithmic recommendations, yet these changes are not undone by 
temporarily down-ranking co-partisan content in one's own feed. 

\item The experiments that change some aspects of the news feed algorithm also clearly violate SUTVA as the treatment outcome depends on the status of others (Spillover, see  Fig.~\ref{fig:sketch}D). Most clearly, when the intervention is to switch to a chronological feed that is still populated with posts that others have received from algorithmic recommendations and re-shared \cite{guess2023social}. To reduce co-partisan content, one study algorithmically down-ranked such content in treated users' feeds \cite{nyhan2023like}. What a user ultimately sees is, however, also dependent on a candidate set of posts shared by their connections, which is implicitly shaped by how algorithmic curation impacts their feeds. 
As the study was only carried out on a small portion of the U.S. population, whatever co-partisan and cross-cutting items participants saw were determined in large part by individuals who were exposed to the unmodified algorithm. As a result, even the individual-level effects in this study are likely biased towards zero by unaddressed SUTVA violations. 

\item More recently a different experiment was conducted on X (previously Twitter), but independently of the platform itself. Here, the researchers directly interfered with the individual feed of participants via a browser add-on that was re-ranking content \cite{piccardi_reranking_2025}. The effects on participants' individual-level metrics of affective polarization (i.e., feelings towards the political out-group) were stronger than in the Meta papers. We interpret them to possibly suffer less from SUTVA violations, because the treatment acted directly on the content level instead of the algorithm or source level. The variable of interest (content expressing antidemocratic attitudes and partisan animosity) was more controlled (in both directions, up- and downward) and could have led to the stronger effects. As a downside, the researchers did not have direct access to the feed algorithm. Their intervention was limited to the up- and down-ranking of content expressing antidemocratic attitudes and partisan animosity. Still, their results show the relative effect to the X-algorithm, but highlighting the typical trade-offs between experimental control and ecological validity, especially without platform-level access. 

\item A Facebook deactivation study by Allcott et al. \cite{allcott2024effects}, by contrast, exhibits inferential challenges from hysteresis. A key outcome of interest was whether Facebook had any impact on election outcomes in the U.S. Yet to the extent that users' voting decisions were influenced by Facebook, opinions on the incumbent candidate (and perhaps the challenger, if well known) would have been formed over years prior to the election and experiment. There is no empirical reason to expect that withdrawal from Facebook has a symmetric effect to being exposed to the algorithm (likely for much longer time periods). 
%The observed point-estimated magnitude of the vote shift being on the scale of what might tip an election (2.6\%) is therefore striking, as even a zero percent change could be entirely consistent with preferences accrued over many years. As the observed vote share change is almost certainly an underestimate of the total causal effect, as in the case of early research on voter turnout described above, the total impact of Meta on elections may well have been appreciable or even determinative. 

\item A more recent field experiment, conducted on X (previously Twitter) \cite{gauthier_political_2026} illustrates the Hysteresis problem very well: In this study there was only an effect detected for the sub-group of participants, who did not previously use the X-Algorithm (they used the chronological timeline instead) and switched to it as the experimental treatment, but no effect for the other way around (i.e., for participants who used the algorithm before and switched to a chronological feed). Those results illustrate very well that algorithmic effects on political opinions and polarization are very likely asymmetrical and that we either need naive populations (which are further diminishing) or different treatments than withdrawal to study them in the future. Still, also this study suffers from SUTVA violations and their effects need to be interpreted with this in mind.

\end{itemize}

Taken together, conducting and interpreting RCTs on social media is challenging due to the complexity inherent in online social networks. And while the authors describe some of these challenges as limitations, they are more accurately viewed as fundamental barriers to meaningful causal inference \cite{munger2025did}. Obviously, this is not to say that such approaches are without value in the study of social media, and we return to their potential later. However, it is important to move past the notion that RCTs inherently and intrinsically produce unbiased causal estimates and acts as a silver bullet to answer causal questions. Instead, the quality of any causal insight---from experiments to observational data analysis---depends wholly on the validity of assumptions underlying the causal inference and not the intrinsic properties of the approach. 

As we have outlined above, the assumptions required by RCTs are by no means guaranteed in the study of social media. Unless the outcome of interest is on the individual level and the treatment independent of others (e.g., attention allocation as a function of linguistic features of a post), estimates of individual-level effects may not provide reliable insight into collective phenomena, and may systematically tend to be under-estimated due to issues such as SUTVA violations. Taken together, claims that small, varying, or null effects from RCTs on the individual level alleviate concerns about broader societal impacts of social media have to be critically examined. 

\section*{From estimated effects to actionable understanding} 

Even if we could conclude with some confidence that collective social-media use negatively impacted some outcome we care about---for example, suppose we found that it increased affective polarization---what would we do with this knowledge? Normally, causal research guides policy-making by helping identify the causes of observed effects towards proposing interventions, or by measuring the relative effect of proposed interventions to facilitate decision-making regarding which policies or changes to adopt. 

If we were to learn that social media use generally increases polarization, it could lead to the insight that decreasing or eliminating social media use might reduce affective polarization. However, social media has become digital public infrastructure with an array of effects, some good and some bad. Its elimination is only conceivable within range of political possibility for specific, vulnerable groups, such as children (a policy decision that indeed was recently made in Australia). Examining, rather than erasing heterogeneity across users, as well as testing feasible interventions are important levers for developing policies that can be applied to address specific harms \cite{Orben2025}. For instance, rather than asking whether or not social media as a whole causes decreased well-being on average, one could instead ask if engagement-based ranking leads to unhealthy overuse among teenagers, if one is concerned about real-word implementation of the findings.  

Examining the heterogeneity of social media effects also extends to different cultural backgrounds. Social media has been taken up by billions of people around the world, yet there still is a heavy bias towards studies from and about the US \cite{lorenz2023systematic}. Given the relevance of these platforms for e.g., politics around the world, the development of actionable interventions must also entail to understand how their effects unfold differently (or similarly) in different cultural background, as it has been done in some existing studies \cite{garimella_images_2020, valenzuela_downward_2022, Bursztyn19}.

It furthermore follows that rather than focus on the net effect, actionable research on social media should seek to identify realistic interventions concerning specific features of platforms that could actually be improved \cite{orben2024mechanisms}. This could include platform interventions, such as bridging \cite{burton_simple_2024, ovadyaBridgingSystemsOpen2023} or down-ranking of toxic content, or user-side interventions such as cross-partisan dialogue \cite{lorenz2020behavioural}. Here it is important to note that another source of heterogeneity comes from different platforms, while collective, system-level outcomes of interests often expand beyond one platform and therefore beyond just one algorithm. Viable and sustainable interventions should hence be evaluated on their systemic impact across multiple platforms, e.g., by cross-platform spillover \cite{mekacher_systemic_2023}.

Although algorithms are certainly an important factor in determining exposure, influential content creators have also disproportionate agency in shaping the content that is posted, and experience unique incentives relative to typical users \cite{lorenz-spreen_2024}. Thus, if interested parties (platforms or policy makers or civil society) want to intervene on exposure to content, an important lever is through creators. To the extent that platform design impacts exposure to content, the incentives that the platform gives creators is an important route by which this impact operates. As a trivial example, demonetizing a particular type of content will probably decrease the rate of production of that type of content. Also, altering the forms of social feedback that are available on a platform, could be a promising avenue to change incentive structures \cite{oswald_collective_2025}.

While it is centrally necessary to run RCTs for testing more targeted, actionable interventions, the  caveats of individual-level RCTs still exist, hence their effects need to be evaluated on the level of collective that are connected on social media.

%Some of the Meta studies discussed earlier \cite{guess2023social, guess2023reshares} actually tested realistic interventions in the form of altered ranking algorithms, but did not assess their collective outcomes at the group-level, which almost certainly lead to under-estimation of their causal potential. Had they been done differently, in particular without violating SUTVA, they may have yielded more valuable outcomes. 

We encourage the research community to open perspectives not only with regards to the collective nature of outcomes but also with regards to policy-relevant treatments. In short, evaluating \textit{actionable interventions} on a group-level has three key benefits over evaluating the effects of social media as a whole: it involves a coherent counterfactual and is thus a conceptually meaningful estimand; it is practically possible to study, because smaller treatments can be put in place for longer than e.g., deactivation studies; and it can guide the design of policy interventions. 

In complex systems, as we described before, hysteresis and other time-dependent phenomena, such as feedback loops, play an important role. It is therefore of importance to note that the effects of interventions will depend on time and/or the state of other variables at that point in time \cite{munger2023temporal}. Consequently, their effects need to be continuously evaluated, as is likely the case for commercial indicators, such as dwell-time and clicks, already\cite{kaufman2017democratizing}. Interventions might be demographically or locality limited and applied for a finite time-period, whereas others may need global and persistent changes. If these interventions aim at collective outcomes, the methods for assessing their causal effects must also account for other properties of complex systems, namely that outcomes are not simply the sum of their parts and that interventions spill over through social networks.

\section*{Embracing epistemic pluralism} 

The challenges of studying social media just described are serious, and our discussion above highlights that progress cannot occur by solely relying on individual-level RCTs, and their findings should not implicitly take precedence over other modes of inference. Fortunately, this is not uncharted territory. Conservation biologists, climate scientists, and public health researchers are rarely able to evaluate their proposed interventions using RCTs at the scale of the whole systems on which interventions are targeted. Nevertheless, they have made substantial progress in developing workable interventions for protecting ecosystems, mitigating climate change, and reducing the large-scale burden of disease. Progress in each of these areas has been heavily reliant on epistemically pluralistic approaches the leverage the strengths and weaknesses of experiments, theory, and observation. 

As with work in climate science and conservation, the search for interventions is often motivated by observational research. Much of the research on social media is similarly motivated by observations that suggest social media is having deleterious societal impacts \cite{Orben2025}. Relevant outcome variables include mental health indicators, beliefs that might undermine democracy, attitudes such as affective polarization, and behaviors such as vaccination, voting, or hate crime. Often these variables exhibit concerning and consistent trends, such as reductions in teen mental health, rising polarization, erosion of democratic norms, and reduced rates of vaccination. So what do we do with these patterns? 

Such observational findings are often dismissed for 
lacking the ``gold standard'' status thought to be associated with RCTs. This is, however, merely an inverse of the misunderstanding about the intrinsic epistemic superiority of RCTs. The assumption that observational studies can solely produce associational or correlational evidence is as fraught as the assumption that RCTs necessarily yield valid causal insights. Both misunderstandings arise from believing that a \textit{method} imbues causality, whereas causality is actually a product of the assumptions we make about the associations we observe in our data both within and across studies \cite{Pearl2000}. 
For that reason, observational data can identify causal relationships to the extent that the assumptions required for identification are credible \cite{angrist1996identification}, which is equally true of experimental data.

While at first glance, it may seem that the discussed limitations may prevent valid estimation of causal effects from social media altogether. However, different approaches come with \textit{different} assumptions. 
If results turn out to be consistent across approaches despite the \textit{variability} in assumptions of, e.g. RCTs and observational studies,
this increases confidence that the effect of interest is not merely an artifact of one particular assumption failure. 
Observational patterns that reveal concerning impacts of
social media, plus theories that suggest plausible mechanisms behind those observations and generate hypotheses for testing,
and finally experiments that test such hypotheses with targeted interventions, in combination will allow to draw a coherent picture of evidence and provide actionable insights for social media research \cite{munafo2018robust}.

\begin{figure}[!htb]
    \centering
    \includegraphics[width=0.9\textwidth]{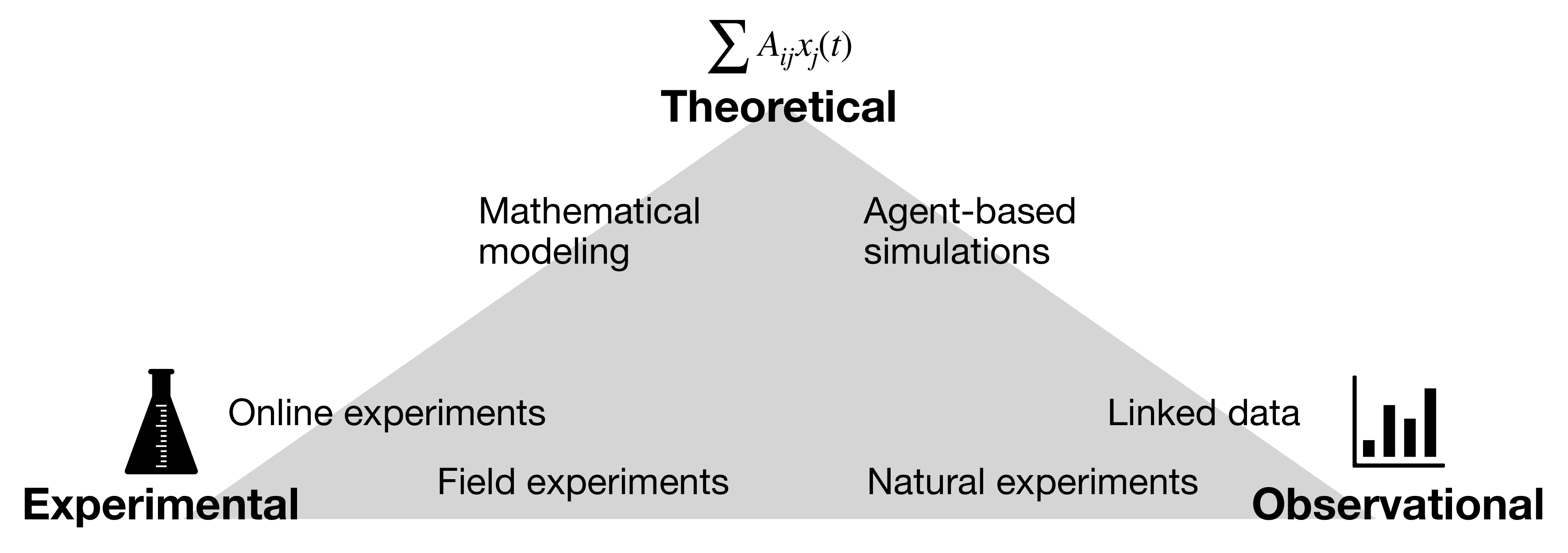}
    \caption{Illustrations of the triangulation approach.}
    \label{fig:triangulation}
\end{figure}

\subsection*{Moving forward}

This perspective allows us to recover the role of RCTs in social media research. For some contexts, questions, and inferential goals they can be powerful tools for insight. Consider common experiments that evaluate changes in mental health or academic performance during a multi-week period of giving up social media.
While this cannot directly tell us about the collective effects of social media over long time-scales, the estimate it produces could certainly inform an individual's decision to take a break.
SUTVA violations here are less of a concern as the person taking a break---as with the participants---remain in a world filled with social media. In other contexts, we may be able to make progress using RCTs and reasonable assumptions about how their effects scale. For example, a design intervention that reduces violent content in one's news feed might support a reasonable inference that applying it broadly would reduce the reach of violent content on a platform. 

Furthermore, while technically difficult and expensive, RCTs that change the unit of randomization from individual to a group-size of interest (e.g., online communities), also referred to as graph cluster-randomization design \cite{ugander_graph_2013, duflo_chapter_2007, aridor_experiments_2024}. Such designs are challenging but can become a powerful tool for answering causal collective questions for social media design and has been employed in controlled, interactive experimental setups \cite{stein2023network}, a collective field experimenton Reddit that made use of the sub-reddit structure \cite{oswald_schulz_lorenz-spreen_2025}, or on the level of municipalities \cite{enriquez_mass_2024}.\footnote{See also ongoing studies on teen mental health that employ such group-level design \url{https://www.cam.ac.uk/stories/irl-trial-social-media-study-launch}}
While this raises the crucial issue of bias-variance tradeoffs and statistical power \cite{aridor_experiments_2024} as the unit of analysis becomes a group, this issue should already be salient with regards to estimates of heterogeneous treatment effect, an estimand of interest for many policy relevant questions \cite{brashier_fighting_2024}.

Ultimately, large scale experiments that run for longer periods of time and with collective-level treatments are necessary. They are without a doubt expensive and difficult to coordinate \cite{munger2025did}, but (1) appropriate given the societal challenges of digitalization, (2) most likely already in use by commercial platforms since decades, also known as via A/B testing \cite{kaufman2017democratizing, ugander_graph_2013}, (3) can be achieved by coordinated efforts, e.g., on the EU-level, and (4) have the potential to turn into collaboratively run citizen-science approaches that allow the population to co-shape alternative platform designs. 

But even with the biggest experiments, a bridge is needed between RCTs that provide insight at the scale of individuals (or groups of participants) in discrete moments (or limited experimental phases), and ``always-on'' \cite{salganik_bit_2019} observational approaches that reveal trends at larger time-scales. 

A line of research addressing causal inference from observational data that may be worth exploring with regards to social media effects is the study of complex climate and ecological systems \cite{sugihara_detecting_2012, runge_detecting_2019}. These approaches extend beyond conventional regression-based methods and the Granger causality framework, which is primarily concerned with predictive relationships among time series \cite{granger_investigating_1969}. In particular, advances in causal discovery for high-dimensional, weakly coupled dynamical systems---conditions that plausibly characterize social media environments---facilitate the identification of causal relationships among multiple interacting variables. A key conceptual foundation of these methods is that relationships between variables arise from their joint embedding within an underlying dynamical system, such that they evolve on a shared attractor manifold.

Convergent cross mapping (CCM), a methodology grounded in this principle, exploits state-space reconstruction to infer causal interactions and is capable of detecting bidirectional coupling, including feedback mechanisms. Intuitively, if one component of a system causally influences another, the historical evolution of the affected component should encode information about the influencing one that can be recovered from its reconstructed state space. This approach has been successfully applied in ecological settings, for instance in the analysis of predator–prey dynamics \cite{sugihara_detecting_2012}. By extension, it is conceivable that such methods may provide a principled framework for identifying causal coupling between, for example, user preferences and algorithmic recommendations in social media time-series data.

More generally, other disciplines tasked with understanding and intervening on complex systems have built such bridges using an empirically-grounded theory- and model-driven paradigm that aims at understanding specific mechanisms. 
Such an approach of bridging experimental, observational and theoretical work (Fig.~\ref{fig:triangulation}), which we feel is necessary for the field to move forward, stands in contrast to relying on single-shot experiments to generate broad statements about platforms' overall impacts. While the community of computational social science captures a broad repertoire of observational, experimental and simulation-based approaches, RCTs have so far attracted most scientific and policy impact.

Although uncommon in the social sciences, the triangulation paradigm is not out of reach. Theory and mechanistic models do not serve to replace empirical research, but rather augment it by bridging gaps across scales of organizational complexity---from individual behavior to societal-level phenomena. If observational approaches to causal inference reveal a trend, theoretical approaches can be used to propose how such trends might arise from mechanisms at play. In turn, the mechanistic assumptions of the multi-scale model can be identified and confirmed at individual scales with formal experiments, providing confidence that the model is able to capture leading-order features governing dynamics across scales. %Computational methods, such as agent-based simulations or interactive online experiments furthermore allow for bridging the methodological gaps.

The spread of misleading information online provides a salient example of how jointly leveraging observational and experimental research can be facilitated by complexity-minded theory. Given observations that misinformation proliferates online, a variety of interventions have been proposed, studied, and experimentally validated in RCTs \cite{kozyreva2024toolbox}. However, despite the success in RCTs and even convincing some platforms to adopt these approaches, estimating their population-level effect is not straightforward as the dynamics of misinformation spreading present a typical example of a complex, non-linear system \cite{Juul21}.
Yet misinformation spreads in a manner that can be captured by models of contagious processes \cite{Juul21}. Beyond mere theory, data can be used to constrain parameters of contagion models, providing a tight fit to observed dynamics and allowing for evaluation of counterfactuals in which various policies are implemented \cite{BakColeman23, BakColeman22} or estimating downstream consequences of misinformation with epidemiological modeling \cite{himelfarb2023fault}. Contagion models can be explicitly made consistent with evidence from RCTs, by hard-coding the observed reductions in sharing behavior from  experiments. 

Using such a model fit to data from the 2020 US elections, one study predicted that despite the nominally large effect size of the ``community notes'' program which relies on appending crowd-sourced fact-checks \cite{wojcik2022birdwatchcrowdwisdombridging}, changes in exposure would be minimal---as large amounts of information dissemination by influential accounts can occur prior to labeling and for unlabeled posts. Observational investigations provided confirmation of these model implications, confirming that community notes were indeed too slow to significantly reduce engagement with misinformation \cite{chuai2023rollout}. In this example, the RCTs are valuable at suggesting an intervention, whereas theory and observation provide insight into its sufficiency at scale. 

These mechanistic models can often be used to evaluate multiple potential interventions in a common framework. Here, for example, the model indicates that dynamics of information spread on platforms will be highly determined by the actions of influential accounts \cite{BakColeman22}. While content creators have disproportionate agency in shaping the content that is created and propagated, which is sometimes cited as an argument against the relative importance of social media platforms, a broader perspective recognizes their dependence of specific incentive structures that are set by the platforms \cite{aridor2024economics}. For example, by using economic models, as content creation on platforms is {\em labor} \cite{duffy2019platform}. And this suggestion that influence incentives provide a uniquely large lever for intervention is indicated on longer time horizons by observational research by Frimer et al., which measured a 23\% increase in incivility among US politicians and attributed it partly to learning from positive feedback for uncivil tweets \cite{frimer2023incivility}. As interventions at the creator level have been comparatively understudied relative to user-specific interventions, this suggests a fruitful avenue for future research. For example, we might ask what can break feedback loops between creators and their audiences that may be driving influencers towards extremism. 

The critical issue is that we can either, as a field, opt to continue scratching our heads over conflicting evidence or attempt to bring it all together. There are risks on both sides: not only the risks of acting on incomplete causal evidence, but also the risks of inaction in the presence of potentially harmful system-level effects. Recognizing large-scale patterns that are consistent across observational approaches which vary in their assumptions can guide us towards where interventions may be needed. RCTs can help us understand proximate, small-scale effects that in turn can fuel theory linking observations and intervention. Expanding them to the collective level with large-scale, citizen-science approaches, will help us understand how those effects are scaling up. Theoretical models, such as variations of agent-based simulations, will help to mechanistically bridge the gap between those micro- and potential macro-findings. In embracing multiple modalities of inference, research on social media can follow in the footsteps of other disciplines that have embraced complexity to develop workable interventions in complex systems (Fig.~\ref{fig:triangulation}). 

Finally, another necessary step for progress and triangulation is some standardization of outcome measures, as in other fields. If we aim to compare experiments, with observational survey data, for example. To be able to interpret the experimental results relative to the survey data, the measures for affective polarization and others, need to be comparable.

\section*{Discussion}
Summaries of social media's impacts have tended to overlook the affordances of observational research while failing to examine the limitations of RCTs---asserting erroneously that evidence from the latter outweighs the former \cite{Budak2024,national2023social}. In doing so, it has been easy to identify concerns yet difficult to reach consensus on societal effects. Actionable interventions at the scale of the problem have remained elusive. 

More generally, it is worth reflecting on our inferential goals when establishing links between platforms and harms. Such quantities are likely to be time-varying, hysteretic, and heterogeneous such that it is unlikely that we could come up with a single estimate of effects on any given domain of concern---much less across domains. Beyond debates about harm, benefits, or lack thereof, the key value in identifying apparent causal links, however tenuous, is a recognition that social media has the potential to impact large-scale phenomena of interest. This, in turn, motivates the search for design principles and regulatory action that can reduce risks and enhance benefits. 

Perhaps the first question we can ask is whether causal evidence is required to warrant intervention. Normative and legal justifications for altering the distribution of content may in many cases be sufficient. Researchers may be able to best support such interventions by simply documenting failures of platforms to prevent the distribution of illegal content of their own accord. The Digital Services Act (DSA) of the European Union is a step towards enabling researchers to do so by providing an access
mechanism for platform data (DSA Article 40). 
Platform interventions could go beyond content moderation and build on our existing understanding of how algorithmic design choice shapes content amplification. At present these choices are decided based on business priorities, and there is little reason they should not, or cannot reflect widely held prosocial goals. Moreover, as the content shared on a platform is disproportionately shaped by the incentives that govern content creation, understanding the behavior and dynamics of content creators is likely to be a particularly productive line of research. 

When firmer causal links across scales of organizational complexity are required, it is important to recognize that no approach in social media research represents a ``gold standard''. As a result, we should not anticipate that any given modality of inference (e.g., RCTs, observation, theory) will be sufficient on its own to identify phenomena and propose interventions. Instead, progress will require coordinated interdisciplinary efforts to scale up RCTs to allow inference on collective and/or heterogeneous estimands while, at the same time, acknowledging the respective paradigmatic limitations we discussed and therefore relying on \textit{epistemic plurality} across lines of evidence with varying assumptions, possibly also leveraging computational and mathematical models to link and infer dynamics across time and scale. Switching to the unit of analysis on the collective level is difficult but increasingly feasible through digitalization itself. While its societal impact is the reason for concerns, digital methods also allow researchers to study those at scale \cite{lazer2009computational}. %For example, by scaling up experiments through online sampling and interactive setups in the browser, or by running field experiments on existing platforms, running large-scale simulations of mechanistic models and by analyzing large amounts of digital, observational data.

We believe this is not only a possible path forward for the field, but a necessary one. The alternative is to endlessly fail to reconcile seemingly conflicting evidence, chasing the ghosts of overlooked assumptions and unanswerable questions. Doing so not only fails to yield actionable insight, but undermines any motivations by stakeholders to deliberate on interventions in the first place. Given this, the community should set aside notions of rote causal impacts and move towards more actionable, epistemically pluralistic, and informative social media research.

\section*{Declarations}

\begin{itemize}
\item Funding.
SL and PLS acknowledges financial support from  the 
Volkswagen Foundation (grant ``Reclaiming individual autonomy and democratic discourse online: How to rebalance human and algorithmic decision making'') and the European Commission (Horizon 2020 grant 101094752 SoMe4Dem). LO was also a beneficiary of Horizon 2020 grant 101094752 (SoMe4Dem).
SL also acknowledges funding from 
the European Research Council
(ERC Advanced Grant 101020961 PRODEMINFO) and the 
Humboldt Foundation through a research award. SL
also receives funding from UK Research and Innovation (through EU Horizon replacement funding grant number 10049415). AO is funded by the Medical Research Council (MC/UU/0030/13 and MR/X028801/1), a UK Research and Innovation Future Leader’s Fellowship (MR/X034925/1) and the Jacobs Foundation. JBC was supported by a grant from Templeton World Charities (TWCF-2023-32581) which supported this research. 

\item Competing interests.
SL has  received funding from Jigsaw (a technology incubator created by Google). SL, PLS and AO also work with the Joint Research Centre of the European Union, both in a paid and unpaid capacity. In the past 36 months, AO has received funding or contracts from the Wellcome Trust, Nathoo Family Trust, Huo Family Foundation, Jacobs Foundation, UK Research and Innovation (incl. Medical Research Council, Economic and Social Research Council and Engineering and Physical Sciences Research Council), the UK Department for Science, Innovation and Technology, National Institute of Health, University of Cambridge, Emmanuel College and St John's College of the University of Cambridge, the Prudence Trust and the Livelihood Impact Fund. She was an unpaid member of the ESRC Smart Data Research UK Programme Board, British Academy Public Policy Committee, UK Department for Education Science Advisory Council, UK Department for Science, Innovation and Technology and UK Department for Culture, Media and Sport College of Experts, Australian eSafety Commissioner Social Media Minimum Age Evaluation Academic Advisory Group, and a paid member of the Digital Futures for Children Centre Advisory Board. JBC has engaged in paid consulting for the United Nations. AN has engaged in paid consulting in litigation regarding social media and adolescent mental health.

\end{itemize}

% Following is a new environment, {scilastnote}, that's defined in the
% preamble and that allows authors to add a reference at the end of the
% list that's not signaled in the text; such references are used in
% *Science* for acknowledgments of funding, help, etc.

% For your review copy (i.e., the file you initially send in for
% evaluation), you can use the {figure} environment and the
% \includegraphics command to stream your figures into the text, placing
% all figures at the end.  For the final, revised manuscript for
% acceptance and production, however, PostScript or other graphics
% should not be streamed into your compliled file.  Instead, set
% captions as simple paragraphs (with a \noindent tag), setting them
% off from the rest of the text with a \clearpage as shown  below, and
% submit figures as separate files according to the Art Department's
% instructions.

\clearpage

% \noindent {\bf Fig. 1.} Please do not use figure environments to set
% up your figures in the final (post-peer-review) draft, do not include graphics in your
% source code, and do not cite figures in the text using \LaTeX\
% \verb+\ref+ commands.  Instead, simply refer to the figure numbers in
% the text per {\it Science\/} style, and include the list of captions at
% the end of the document, coded as ordinary paragraphs as shown in the
% \texttt{scifile.tex} template file.  Your actual figure files should
% be submitted separately.

\end{document}